%% file: Continuous_OCT_VF.tex
\documentclass[a4paper, amsfonts, amssymb, amsmath, reprint, showkeys, nofootinbib, twoside]{revtex4-1}
\usepackage[english]{babel}
\usepackage[utf8]{inputenc}
\usepackage[colorinlistoftodos, color=green!40, prependcaption]{todonotes}
\input{preamble}
\usepackage[pdftex, pdftitle={Article}, pdfauthor={Author}]{hyperref} % For hyperlinks in the PDF
\begin{document}

%\title{Continuous FF-OCT for bio-imaging applications}
\title{Continuous scanning full-field OCT for fast volumetric imaging of multi-cellular aggregates}

\author{Charlie Kersuzan}
    \email[Correspondence email address: ]{charlie.kersuzan@u-bordeaux.fr}% Your name
    \affiliation{LP2N, Laboratoire Photonique Numéerique et Nanosciences, Univ. Bordeaux, F-33400 Talence, France}
    \affiliation{Institut d'Optique Graduate School $\&$ CNRS UMR 5298, F-33400 Talence, France}

\author{Anirban Jana}
    \affiliation{LP2N, Laboratoire Photonique Numéerique et Nanosciences, Univ. Bordeaux, F-33400 Talence, France}
    \affiliation{Institut d'Optique Graduate School $\&$ CNRS UMR 5298, F-33400 Talence, France}
    \affiliation{Treefrog Therapeutics, Pessac, France}

\author{Amaury Badon}
    \affiliation{LP2N, Laboratoire Photonique Numéerique et Nanosciences, Univ. Bordeaux, F-33400 Talence, France}
    \affiliation{Institut d'Optique Graduate School $\&$ CNRS UMR 5298, F-33400 Talence, France}

\email{\authormark{*}charlie.kersuzan@u-bordeaux.fr} %% email address is required; see note below about the corresponding author designation

% use {asbstract*} to suppress the copyright line. Copyright information will be added in production

\begin{abstract} 
Full-field optical coherence tomography (FF-OCT) offers label-free, high-resolution imaging of biological samples but remains limited by slow acquisition due to piezoelectric mirror modulation. We present a continuous-scanning FF-OCT method that eliminates piezoelectric displacement and synchronization by continuously translating the sample with a motorized stage while recording images on the fly. Depth-resolved information is retrieved via Fourier analysis of the temporal signal at each pixel. This approach enables volumetric imaging over several hundred micrometers within tens of seconds and provides a higher contrast-to-noise ratio than traditional four-phase FF-OCT. Continuous-scanning FF-OCT thus represents a simpler and faster alternative for 3D bio-imaging of living tissues and organoids.

\end{abstract}

\maketitle

%%%%%%%%%%%%%%%%%%%%%%%%%%  body  %%%%%%%%%%%%%%%%%%%%%%%%%%
\section{Introduction}

Optical microscopy provides micrometer-scale resolution for the observation of biological specimens across spatial ranges of several hundred microns, while maintaining relatively low invasiveness \cite{murphy2012fundamentals}. Several microscopy techniques—such as optical coherence tomography (OCT) \cite{huang1991optical}, light-sheet microscopy \cite{stelzer2021light}, confocal microscopy \cite{wilson1990confocal}, and light-field microscopy \cite{levoy2006light}, are especially effective for observing living samples in 3D, a task that is inherently complex. Among all these techniques, OCT stands out due to its label-free nature and its ability to image deep within tissues \cite{drexler2008optical}. In particular, full-field optical coherence tomography (FF-OCT) has recently emerged as an excellent tool for bio-imaging applications, allowing both structural imaging \cite{beaurepaire1998full,dubois2002high} as well as specific imaging of live cells with recent developments of dynamic OCT \cite{apelian_dynamic_2016}. With an improved lateral resolution compared to conventional OCT, this approach allows precise imaging of skin samples \cite{latrive2011vivo}, eyes \cite{scholler_probing_2019}, mice embryos \cite{morawiec_full-field_2024}, and more recently organoids \cite{scholler_dynamic_2020}: cellular aggregates of cells of several hundred microns in diameter that can recapitulate the 3D architecture and function of various living tissues.

Despite recent progress aimed at improving sensitivity \cite{monfort2025rolling}, field of view \cite{recher2020remote}, or imaging/optical sectioning depth \cite{auksorius2022spatio}, FF-OCT remains partly limited by its acquisition rate. This results from its underlying principle, which we explain in detail below. In FF-OCT, light from a source is split into two channels using a beam splitter. The light reflected by the sample in one arm and by a reference mirror in the other arm is then recombined, and the resulting interferences are detected with a camera. Traditionally, several intensity images are captured for several positions of the reference mirror displaced with a piezoelectric actuator. Typically, the linear combination of four images provides a signal directly proportional to the reflectivity of the sample at a given depth. The acquisition of a 3D volume therefore requires repeating this synchronized procedure for different axial positions of the sample, which is time-consuming and thus limits this method.

In this paper, we propose an alternative method to acquire FF-OCT images without the need of any piezoelectric displacement and synchronization. Here we implemented a continuous displacement of the sample with a motorized stage while acquiring 2D en-face images on-the-fly. Then, for each pixel, a Fourier transform is applied to the time-varying intensity which efficiently extracts the displacement information. This approach, that we called continuous scanning FF-OCT, allows us to acquire complete 3D volumes of several hundred micrometers in depth in a tens of seconds (typically 400 µm in 100 seconds). Since no piezoelectric displacement is required, there is no need for synchronization between displacement and frame acquisition, resulting in faster acquisition times and simplicity. A comparison between our method and a four phase-step traditional FF-OCT showed that a better contrast-to-noise ratio (CNR) is even obtained with the continuous scanning FF-OCT.

\begin{figure*}[ht!]
    \centering
    \includegraphics[width=0.9\textwidth]{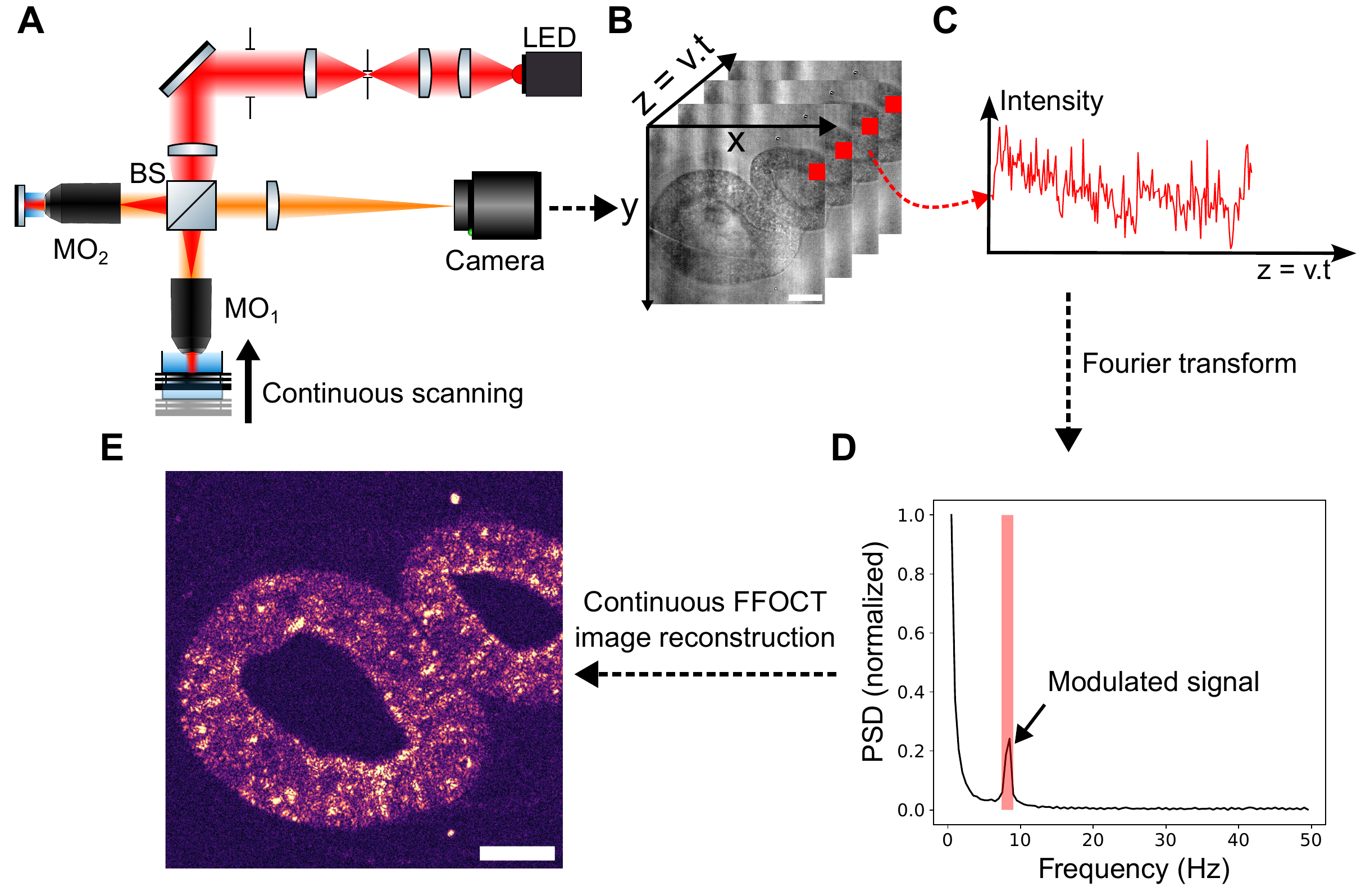}
    \caption{Continuous scanning FF-OCT principle. (A) Experimental setup. The sample placed in the object arm is moved axially at a constant speed using a motorized stage. BS, Beam Splitter; MO, microscope objective; LED, light emitting diode.    
    (B) Time series of raw images of a living multi-cellular aggregate of pluripotent stem cells. The temporal evolution of each pixel is processed independently. (C) Temporal evolution of the intensity of the pixel indicated by a red square in (B). 
    (D) Power spectral density of the same pixel. The red band indicates the target frequency of 8.5 Hz corresponding to a motor speed of 2 µm/s and an acquisition rate of 100 frames per second. (E) Continuous scanning FF-OCT image obtained by repeating this operation for each pixel. Scale bar, 50 µm. Images acquired using 10X objectives.}
    \label{fig:figure_1}
\end{figure*}

\section{Results}

By combining low-photodamage and optical sectioning, FF-OCT can characterize the shape and the structure of multi-cellular aggregates over the course of their development, typically several days. Here, these samples are obtained with a microfluidics technique called the cellular capsule technology (CCT) that produces at high throughput ($\sim$ 5000 per second) suspensions of human-induced pluripotent stem cells (hiPSCs) and matrigel inside a porous membrane of alginate \cite{alessandri_cellular_2013}. These so-called capsules are around 400 to 600 µm in diameter and the cells inside grow over time in 3D, forming cysts: a monolayer of hiPSCs with a lumen inside \cite{cohen2023engineering}. In this work, we present images of empty capsules as well as capsules containing cysts, alive and fixed.

\subsection{Experimental set-up}

Figure \ref{fig:figure_1}(A) shows a schematic of the experimental setup. A broadband and spatially incoherent light source centered at 625 nm (M625L4, Thorlabs) illuminates a Michelson interferometer in a Kohler configuration made of four lenses and two apertures. The light source was chosen for its spatial incoherence and high power, enabling high signal-to-noise OCT imaging near camera saturation. A wavelength of 625 nm was selected for its low phototoxicity, deep penetration, and compatibility with visible-range optics and the camera. The source coherence length, defined as the full width at half maximum of the axial fringe pattern, is about 11 µm and determines the axial resolution of the microscope. Light is then separated by a beam splitter (CCM1-BS013, Thorlabs) and propagates through two identical microscope objectives (MO). Depending on the experiments, two different pairs of water immersion objectives are used: 10X (UMPLFLN 10X NA=0.3, Olympus) and 20X (UMPLFLN 20X NA=0.5, Olympus). The specimen of interest is placed in the focal plane of one of the MO onto a motorized stage (PI M231.17) in order to allow high precision continuous sample displacement. In the second arm, a silicon mirror is placed and supported by a piezoelectric transducer (PK25FA2P2, Thorlabs) modulated by a data acquisition card (BNC-2110, National Instruments) and a controller (MDT694B, Thorlabs). Light reflected in the two arms is then collected by the same MO, focused by a 200 mm tube lens (Thorlabs TTL200-A) and imaged by a CMOS camera (MV1-D1024E-160-CL-12, PhotonFocus). A field of view of 972 x 972 µm$^2$ and 486 x 486 µm$^2$ is obtained with the 10X and 20X objectives respectively. Additionally, a manual stage (PT1, Thorlabs) positioned in the reference arm between the beam splitter and the objective  allows the tuning of the optical path length to match the coherence plane with the focal plane of the objectives. 

\subsection{Image acquisition and reconstruction process}

The continuous scanning OCT 3D image reconstruction method is represented in figure \ref{fig:figure_1}. The specimen of interest is continuously moved axially at a speed of a few µm/s while acquiring images continuously with a frame rate of typically 100 Hz. Typically, a stack of 10 000 images covers an entire volume of 400 µm in 100 seconds at a speed of 4 µm/s. The displacement of the sample results in a fluctuation of the interferometric term, which in turn leads to fluctuations in the detected intensity (see figure \ref{fig:figure_1}(B)). This behaviour is expected, since the detected intensity at a given time is governed by the following equation:
\begin{equation}
    I = I_0\left(R_r + R_{\text{sample}} + 2 \sqrt{R_r R_{\text{sample}}}\cdot\cos\Phi         \right)
\end{equation}
where $R_r$ is the reflection coefficient of the reference mirror, $R_{\text{sample}}$ is the reflection coefficient of the object in the interference plane (the signal of interest) and $\Phi$ is the phase difference between the two arms (see SI for more details). The frequency of the fluctuations of the interference signal is directly proportional to the motor speed, as illustrated in figure \ref{fig:figure_1}(C). Thus, it is crucial that the motor displacement speed remains constant in order to maximize the signal associated with this frequency, while also preserving a proportional relationship between time and depth ($z = vt$). Once an entire volume is acquired, the data processing consists of the following steps for each sub-stack of 100-200 images :

\begin{itemize}
    \item First, for each pixel, a Fast Fourier Transform (FFT) is computed along time in order to retrieve the power spectral density (PSD). To estimate the position of the target frequency, the average PSD of the entire image is first computed. A frequency component is identified as the target if its value exceeds 15\% of the PSD at zero frequency. All neighboring components with values greater than 15\% of the PSD at the target frequency are included as part of the signal of interest. This method compensates for motor speed fluctuations across frames.

    \item Then, for each pixel, the intensity of the PSD considered as part of the target frequency are summed together and assigned to that pixel as the Continuous FF-OCT signal.
    
    \item The resulting image is normalized between 0 and 1. Last, to compensate for frame to frame instability due to variations in stage speed, the top 0.5\% of pixel intensities are saturated, while the bottom 0.5\% are set to zero.    
\end{itemize}

Figure \ref{fig:figure_1}(E) shows the resulting image of a living multi-cellular aggregate obtained with this method for a motor speed of 2 µm/s  and with 200 frames acquired over 2 seconds. The structure of two monolayer cysts inside the alginate capsule are clearly observable. The entire processing takes around 1 second, with a computer composed of an Intel Xeon W-2145 @3.70GHz CPU, an Nvidia Quadro P2000 GPU and 64 Gb of RAM. Data processing is carried out in Python, using CuPy to perform the entire image processing on the GPU.

\subsection{Effect of stage speed on image quality}

In the previous experiment, the motor speed was set to 2 µm/s. We now examine how this parameter influences image quality. To quantify image quality, we measure the contrast-to-noise ratio (CNR), defined as the difference between the average signal of interest and the background signal, divided by the standard deviation of the background, and expressed in dB:

\begin{equation}
    CNR = 20 \, \text{log} \left(\frac{I_{\text{foreground}}-I_{\text{background}}}{\sigma_{\text{background}}}\right)
\end{equation}

In the rest of the paper, the signal of interest is estimated inside a squared region corresponding to the cellular aggregates, while the background signal is estimated in a squared region corresponding to a dark region of the image. We investigated the effect of the motor speed on the image quality, with speeds ranging from 0.5 to 5 µm/s, by imaging the same sample as previously at each speed, acquiring 200 images at 100 fps. 

\begin{figure*}[ht!]
    \centering
    \includegraphics[width = 0.9\textwidth]{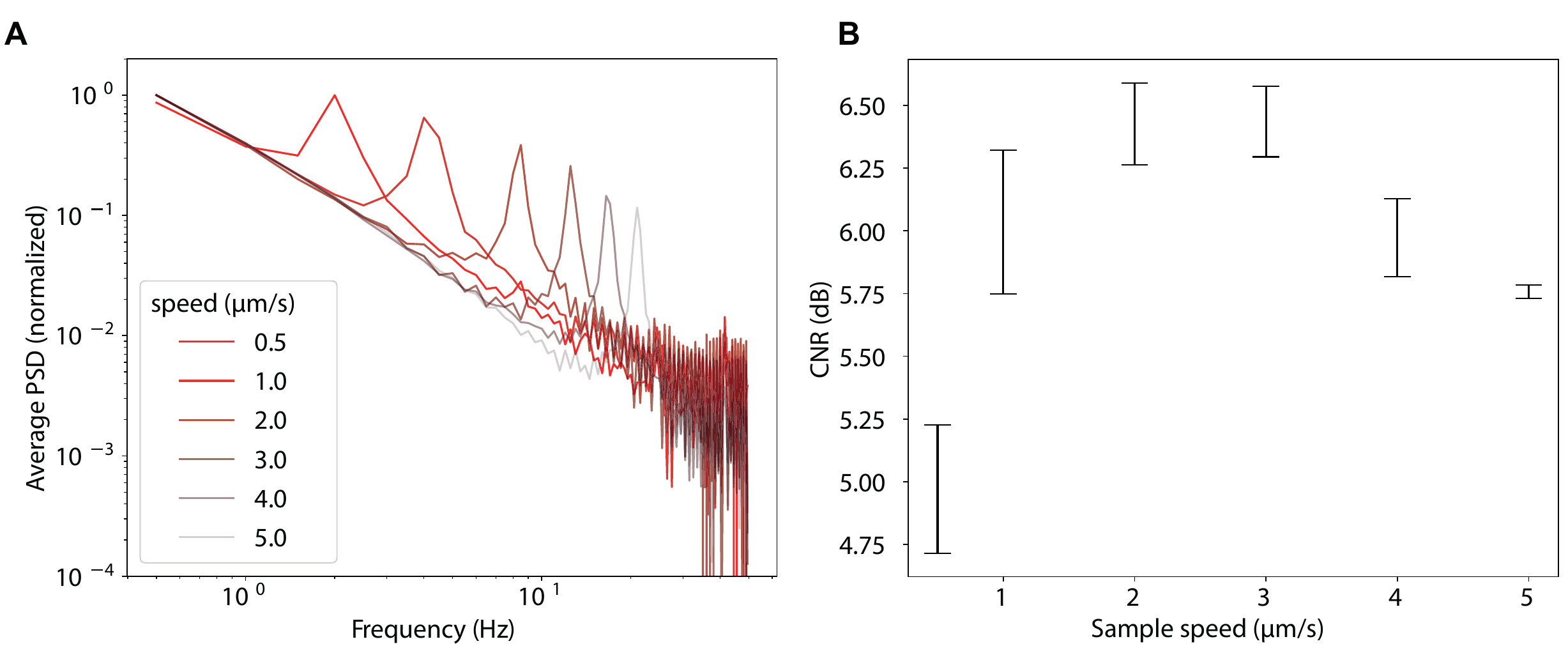}
    \caption{
    Effect of stage speed on the performance of the continuous scanning FF-OCT approach. (A) Average power spectral density obtained over the entire image for the same displayed in figure 1 scanned through at different motor speeds. Images for the different speeds are presented in supplementary materials. (B) Contrast-to-noise ratio (CNR) obtained at different sample speeds. In each case, the camera frame rate and the number of images are kept constant.}
    \label{fig:figure_2}
\end{figure*}

Figure \ref{fig:figure_2}(A) represents averaged PSDs over the entire image plotted in log scale for different motor speeds at the same $z$ plane within the sample. As expected, the target frequency evolves linearly with the speed of the motor, indicating the linear behaviour of the motor employed in this study. We then compared the CNR obtained for the same experimental conditions in the continuous scanning FF-OCT frames (see \ref{fig:figure_2}(B) and SI for the images). At a frame rate of 100 fps, the contrast-to-noise ratio (CNR) reaches its maximum for motor speeds of 2–3 µm/s and exhibits a significant reduction at both lower and higher velocities. For each regime, this quality reduction can be attributed to distinct underlying mechanisms. At higher speeds, the reduction in CNR results from fewer images being acquired over a 2$\pi$ phase shift, thus approaching the Nyquist limit and lowering the effective sampling rate at the target frequency. At slower speeds, the reduction in CNR is instead caused by the presence of a signal at $f = 0$ Hz. This originates from the constant component generated by the strong light reflected from the reference mirror. When the signal fluctuates at frequencies close to 0 Hz, it becomes challenging to distinguish the background noise from the signal of interest.
It should be noted that these results were obtained using the same number of frames in all cases. Increasing the number of frames at a given depth by acquiring frames over a larger z-volume increases the CNR, though at the expense of z-axis resolution (see Fig. S2): in order to satisfy the Nyquist criterion, the volume used to calculate the FFT needs to be at least twice smaller than the coherence length of the light source (approximately 5 µm in our case). We also examined the influence of the temporal window on CNR and found that longer windows yield better performance than summing shorter ones (see Fig. S3). 

For the remainder of this study,the motor speed is set to 2 µm/s and the data processing was performed on sub-stacks of 100 frames, this number being chosen to ensure a window size approximately four times smaller than the coherence length of the illumination source. 

\subsection{Comparison between 4-phase and Continuous scanning}

\begin{figure*}[ht!]
    \centering
    \includegraphics[width = 0.9\textwidth]{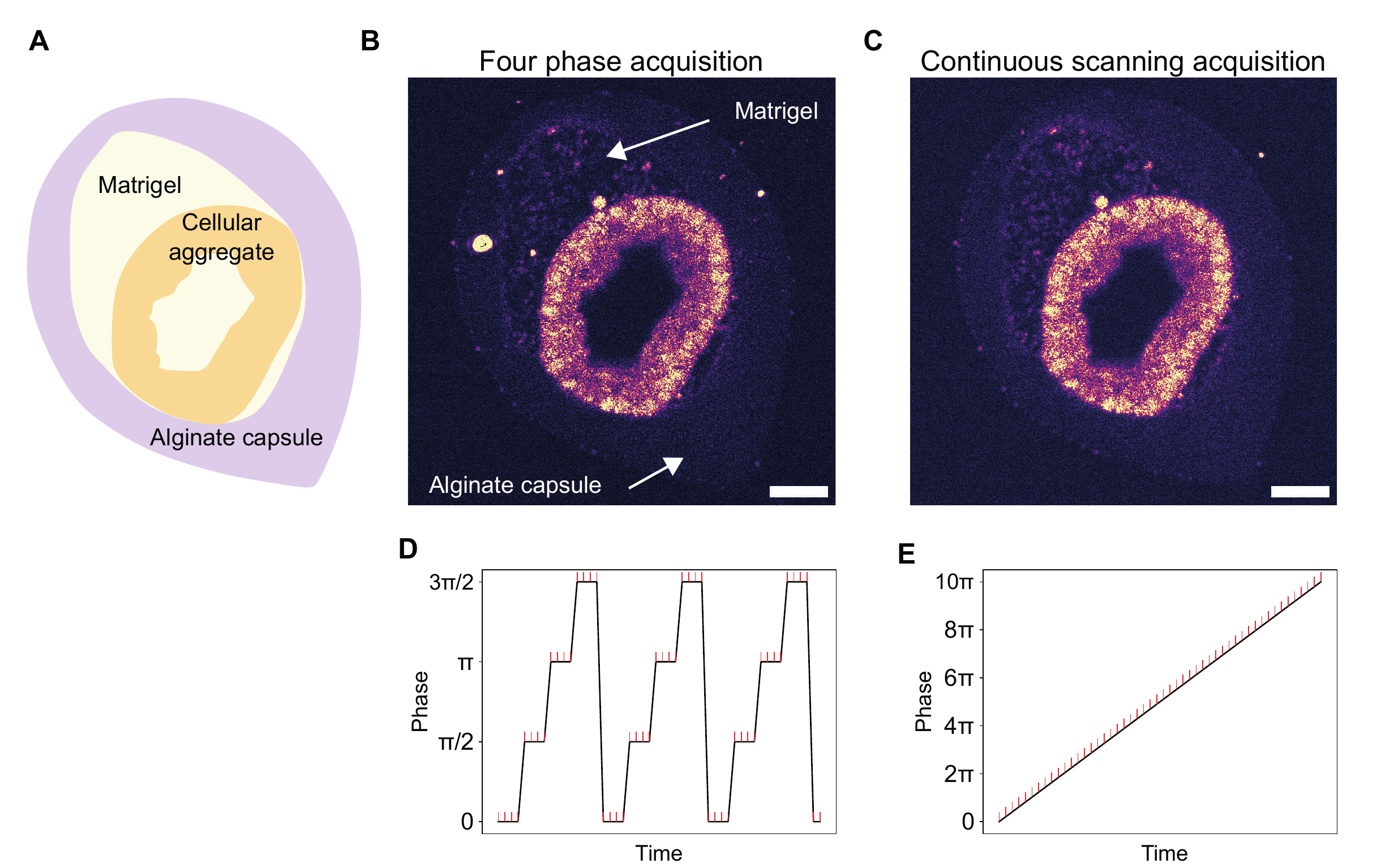}
    \caption{Comparison between the four-phase acquisition method and continuous scanning mode in FF-OCT of a living multi-cellular aggregate. (A) Schematic view of an alginate capsule filled with matrigel within which a multicellular aggregate is growing. (B) FF-OCT image obtained by modulating the reference mirror with a piezoelectric actuator in a four-phase scheme and reconstructed with 96 images. (C) Continuous scanning FF-OCT image of the same sample at the same z-plane reconstructed with 100 images at 2 µm/s.
    (D) Principle of four-phase step acquisition. The piezoelectric displaces the reference mirror in order to acquire images with specific phase steps and averages over them.
    (E) Principle of continuous scanning acquisition. The sample is continuously moved through the interference plane and the phase of a specific plane varies continuously during acquisition.
    Scale bars, 50 µm. Images were acquired using 20X MO.}
    \label{fig:figure_3}
\end{figure*}

To assess the imaging quality of our approach relative to conventional FF-OCT methods, we compared images acquired using the traditional four-phase FF-OCT algorithm with those obtained through continuous-scanning FF-OCT with the parameters set previously. For the four-phase acquisition, at each piezoelectric position corresponding to a phase shift (0, $\pi$/2, $\pi$, and 3$\pi$/2), 12 images are captured and averaged, and this procedure is repeated twice. Figure \ref{fig:figure_3}(B) and (C) display images obtained with a traditional four-phase FF-OCT algorithm (N=96 images) and continuous FF-OCT algorithm (N=100 images). The two images show almost identical imaging qualities. Measuring the CNRs of the two images, we obtain values of respectively 16.4 and 17.6 dB, showing a non-negligible gain of 1.2 dB in CNR for the continuous FF-OCT image. Interestingly, we notice in the continuous FF-OCT image a substantial reduction in bright spots associated with the presence of dust or tiny particles suspended and moving within the culture medium. If the motion of the particle is larger than a pixel on an image during the four-phase acquisition, it will appear bright while the FFT analysis on a per-pixel basis 
will reject or limit this signal.

In addition to its superior image quality, our method also offers faster acquisition times and greater simplicity. Unlike the four-phase method, which requires synchronization between the camera and the piezoelectric actuator, which involves camera pauses to allow the actuator to reach the correct position, our approach requires no synchronization, enabling the camera to operate at its maximum frame rate without any pauses. While acquisition time in the four-phase method can be reduced through optimized synchronization between frame capture and the piezoelectric actuator, we observed a 2.5-fold decrease in acquisition time with our method (1 s vs 2.5 s) for 100 frames.

Finally, it takes only 0.5 seconds to process a sub-stack of 100 raw images to obtain a continuous FF-OCT image, making it compatible for live imaging at a moderate frame rate, typically 1 fps.

\subsection{Volumetric imaging}

\begin{figure*}[ht!]
    \centering
    \includegraphics[width = 0.9\textwidth]{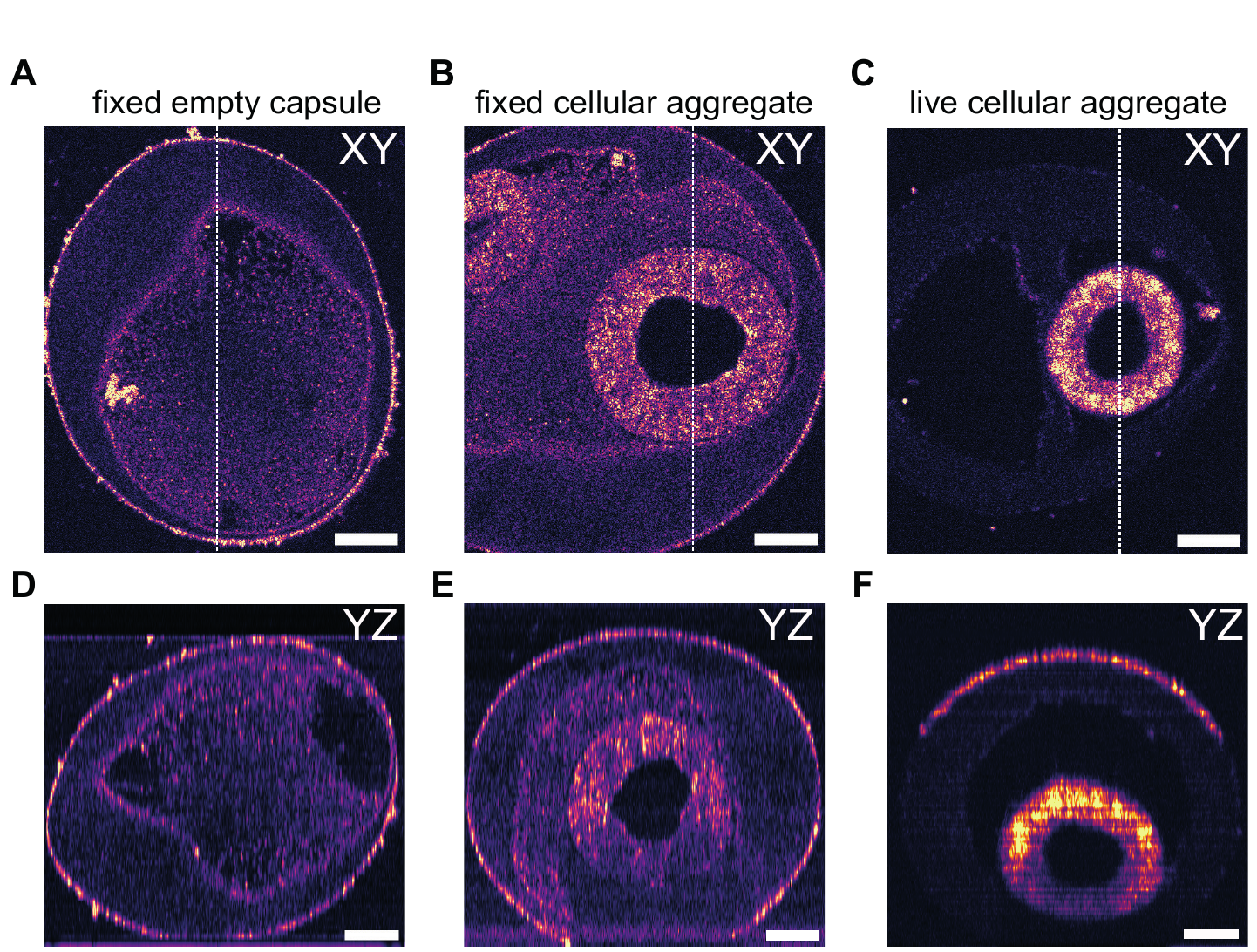}
    \caption{Volumetric imaging with continuous scanning FF-OCT. (A) Continuous FF-OCT image of a fixed cellular aggregate encapsulated in an alginate sphere where the dashed line represents the plane where the transversal view represented in (D) was taken.
    (B) Continuous FF-OCT image of a fixed empty capsule of alginate where the dashed line represents the plane where the transversal view represented in (E) was taken.
    (C) Continuous FF-OCT image of a living cellular aggregate encapsulated in an alginate sphere where the dashed line represents the plane where the transversal view represented in (F) was taken. Scale bars are 50 µm. Images were acquired using the 10X objectives.
    }
    \label{fig:figure_4}
\end{figure*}

At last, our technique was applied to the volumetric imaging of entire 3D samples: an empty alginate capsule, a fixed, and a live multi-cellular aggregate in an alginate capsule. In all cases, the experimental conditions were kept identical, with acquisitions conducted over 200 s, corresponding to an axial displacement of 400 µm at a speed of 2 µm/s.

First, figure \ref{fig:figure_4}(A) and (D) represent respectively an en-face and a transverse cross-section of a fixed capsule filled with matrigel. These findings demonstrate the capability to rapidly and precisely visualize the external and internal 3D structure of a capsule. Notably, the matrigel is observed to be inhomogeneously distributed during production via the CCT method. Such insights are valuable for capsule metrology and for optimizing production conditions to achieve greater uniformity \cite{suire2025extending}. We then imaged a fixed multi-cellular aggregate of hiPSCs. The resulting en-face and transverse cross-section images are displayed in figure \ref{fig:figure_4}(B) and (E). Note that the sample size surpasses the field of view (here 486 µm), and consequently, only part of the sample can be observed. A satisfactory contrast is obtained throughout the sample, allowing clear visualization of the cell monolayer and the matrigel within the alginate capsule. Finally, the same acquisition was performed on a live multi-cellular aggregate. As seen in figure \ref{fig:figure_4}(C) and (F), significantly improved contrast is obtained for the cell aggregate, enabling optimal visualization of its 3D structure.

These results demonstrate the potential of our method for 3D imaging of samples exhibiting relatively low refractive index differences.
Upon fixation of the samples with paraformaldehyde, a decrease of refractive index between 0.002 and 0.1 was previously measured on mammalian cells \cite{baczewska_refractive_2021}. Since the OCT signal is sensitive to refractive index differences, tissue fixation reduces these differences and consequently the signal originating from the aggregate. Thus, without fixation, the signal from the aggregate is significantly improved, allowing for a finer observation of cellular details, although this occurs at the expense of the signal from the matrigel. A precise quantification of the refractive index variations induced by the fixation of the samples, both on the hiPSCs cells and on the alginate capsules, would provide valuable information to confirm this hypothesis.

\section{Discussion}

The continuous scanning FF-OCT technique we developed in this work provides a simple, label-free, fast and efficient imaging technique to observe 3D biological samples at a micrometer-scale resolution. Compared to the conventional FF-OCT technique with a four-phase acquisition, our approach eliminates the need for a piezoelectric actuator to modulate the position of the mirror in the reference arm. In addition to simplifying the experimental setup and reducing costs, the modification eliminates the need for synchronization between the camera and the piezoelectric actuator, which results in faster acquisition times. In our case, acquisition with continuous-scanning FF-OCT was approximately 2.5 times faster, although further optimization of the four-phase method could reduce this gap. In this new configuration, acquisition speed is only limited by the camera frame rate, which is 100 fps in our case. Using cameras with a higher frame rate, the stage speed could be easily increased to enable faster volumetric imaging, as long as a sufficient number of images are acquired over a 2$\pi$ phase shift (corresponding to a sample displacement of $\sim$ 600 nm) to satisfy the Nyquist criterion. 

The proposed approach involves selecting a sub-stack of the entire dataset (N=100 frames) and applying a FFT along the temporal dimension to isolate the target frequency of the sample motion. This operation enables the removal of a significant fraction of noise and non-interferometric signal components, similarly to what was previously demonstrated on a 2 $\pi$ modulation only \cite{morawiec2024full}. Continuous scanning FF-OCT yields images with a 1.2 dB higher CNR than traditional four-phase FF-OCT reconstructed from an equivalent number of raw frames. This provided sufficient contrast to observe multicellular aggregates and the surrounding matrigel encapsulated in an alginate capsule up to 400 µm deep. Within live cellular aggregates, we can detect variations in refractive index that may correspond to different cellular components, such as the nucleus. Parallel fluorescence imaging could help correlate these refractive index differences with specific cell structures. In the future, we aim to develop a dual-acquisition system to capture both fluorescence and FF-OCT images simultaneously.

Our approach shifts the complexity from the experimental setup to the data processing stage. Hence, it is crucial to verify that the reduction in acquisition time is not lost during data processing. In practice, the reconstruction process to obtain a single image from 100 raw frames takes 0.5 seconds on a regular computer equipped with a graphic card (Nvidia Quadro P2000) while image acquisition is performed by a frame grabber (PCIe 1433, NI) and transferred from the RAM memory to the SSD memory. A substantial time gain could be realized by transferring data directly from RAM to the GPU, which would allow real-time FF-OCT image rendering. Such a time gain would then allow the implementation of more advanced analysis methods such as the Welch method, now commonly used in dynamic FF-OCT to extract fluctuations in OCT signal due to intracellular motion. Compared to the FFT, the Welch method is slower but provides results less sensitive to noise. In addition to being more robust, such an analysis would also allow the deduction of a dynamic contrast by extracting from the PSD the frequencies corresponding to the signal generated by intracellular activity. 
A combination of structural and dynamic FF-OCT combined together was recently achieved at a specific plane of the sample by active phase modulation of the sample using a vibrating piezoelectric at a specific target frequency with an amplitude of a few nanometers \cite{yin_fast_2025}.
In the future, we plan to obtain similar results with continuous-scanning FF-OCT.

\section{Conclusion}

In this work, we proposed a new acquisition scheme in FF-OCT to perform volumetric imaging in a simplified experimental setup, by continuously moving the sample in the axial direction and capturing images on the fly. We demonstrated the benefits of our approach both in terms of image quality, acquisition speed, and simplicity of the experimental setup. This last point in particular makes it possible to develop simpler, more compact FF-OCT systems that could be integrated into an incubator to monitor the long-term development of multi-cellular aggregates over several days or weeks \cite{badon2022incubascope, jana2025zincubascope}.
\bigskip

\textbf{Funding}

The authors acknowledge the financial support from the French National Agency for Research (ANR-22-CE42-0019), Idex Bordeaux (Research Program GPR Light), the Interdisciplinary and Exploratory Research grant from Bordeaux University. This work has benefited from a government grant managed by the Agence Nationale de la Recherche under the France 2030 program, reference ANR-24-EXME-0002.
\bigskip

\textbf{Acknowledgment}

We would like to thank all the BiOf team for fruitful discussions and especially Elsa Mazari-Arrighi, Aurélien Richard and Adeline Boyreau for providing biological samples.
\bigskip

\textbf{Disclosures}

The authors declare no conflicts of interest.
\bigskip

\textbf{Data Availability Statement}

Data underlying the results presented in this paper can be obtained from the authors upon reasonable request.
\bigskip

%%%%%%%%%%%%%%%%%%%%%%% References %%%%%%%%%%%%%%%%%%%%%%%%%

%%%%%%%%%% If using BibTeX:
\bibliography{sample}

\end{document}

%% file: preamble.tex
\usepackage{amsthm}
\usepackage{mathtools}
\usepackage{physics}
\usepackage{xcolor}
\usepackage{graphicx}
\usepackage[left=23mm,right=13mm,top=35mm,columnsep=15pt]{geometry} 
\usepackage{adjustbox}
\usepackage{placeins}
\usepackage[T1]{fontenc}
\usepackage{lipsum}
\usepackage{csquotes}

%% file: Continuous_OCT_VF.bbl
\begin{thebibliography}{24}%
\makeatletter
\providecommand \@ifxundefined [1]{%
 \@ifx{#1\undefined}
}%
\providecommand \@ifnum [1]{%
 \ifnum #1\expandafter \@firstoftwo
 \else \expandafter \@secondoftwo
 \fi
}%
\providecommand \@ifx [1]{%
 \ifx #1\expandafter \@firstoftwo
 \else \expandafter \@secondoftwo
 \fi
}%
\providecommand \natexlab [1]{#1}%
\providecommand \enquote  [1]{``#1''}%
\providecommand \bibnamefont  [1]{#1}%
\providecommand \bibfnamefont [1]{#1}%
\providecommand \citenamefont [1]{#1}%
\providecommand \href@noop [0]{\@secondoftwo}%
\providecommand \href [0]{\begingroup \@sanitize@url \@href}%
\providecommand \@href[1]{\@@startlink{#1}\@@href}%
\providecommand \@@href[1]{\endgroup#1\@@endlink}%
\providecommand \@sanitize@url [0]{\catcode `\\12\catcode `\$12\catcode `\&12\catcode `\#12\catcode `\^12\catcode `\_12\catcode `\%12\relax}%
\providecommand \@@startlink[1]{}%
\providecommand \@@endlink[0]{}%
\providecommand \url  [0]{\begingroup\@sanitize@url \@url }%
\providecommand \@url [1]{\endgroup\@href {#1}{\urlprefix }}%
\providecommand \urlprefix  [0]{URL }%
\providecommand \Eprint [0]{\href }%
\providecommand \doibase [0]{http://dx.doi.org/}%
\providecommand \selectlanguage [0]{\@gobble}%
\providecommand \bibinfo  [0]{\@secondoftwo}%
\providecommand \bibfield  [0]{\@secondoftwo}%
\providecommand \translation [1]{[#1]}%
\providecommand \BibitemOpen [0]{}%
\providecommand \bibitemStop [0]{}%
\providecommand \bibitemNoStop [0]{.\EOS\space}%
\providecommand \EOS [0]{\spacefactor3000\relax}%
\providecommand \BibitemShut  [1]{\csname bibitem#1\endcsname}%
\let\auto@bib@innerbib\@empty
%</preamble>
\bibitem [{\citenamefont {Murphy}\ and\ \citenamefont {Davidson}(2012)}]{murphy2012fundamentals}%
  \BibitemOpen
  \bibfield  {author} {\bibinfo {author} {\bibfnamefont {D.~B.}\ \bibnamefont {Murphy}}\ and\ \bibinfo {author} {\bibfnamefont {M.~W.}\ \bibnamefont {Davidson}},\ }\href@noop {} {\emph {\bibinfo {title} {Fundamentals of light microscopy and electronic imaging}}}\ (\bibinfo  {publisher} {John Wiley \& Sons},\ \bibinfo {year} {2012})\BibitemShut {NoStop}%
\bibitem [{\citenamefont {Huang}\ \emph {et~al.}(1991)\citenamefont {Huang}, \citenamefont {Swanson}, \citenamefont {Lin}, \citenamefont {Schuman}, \citenamefont {Stinson}, \citenamefont {Chang}, \citenamefont {Hee}, \citenamefont {Flotte}, \citenamefont {Gregory}, \citenamefont {Puliafito} \emph {et~al.}}]{huang1991optical}%
  \BibitemOpen
  \bibfield  {author} {\bibinfo {author} {\bibfnamefont {D.}~\bibnamefont {Huang}}, \bibinfo {author} {\bibfnamefont {E.~A.}\ \bibnamefont {Swanson}}, \bibinfo {author} {\bibfnamefont {C.~P.}\ \bibnamefont {Lin}}, \bibinfo {author} {\bibfnamefont {J.~S.}\ \bibnamefont {Schuman}}, \bibinfo {author} {\bibfnamefont {W.~G.}\ \bibnamefont {Stinson}}, \bibinfo {author} {\bibfnamefont {W.}~\bibnamefont {Chang}}, \bibinfo {author} {\bibfnamefont {M.~R.}\ \bibnamefont {Hee}}, \bibinfo {author} {\bibfnamefont {T.}~\bibnamefont {Flotte}}, \bibinfo {author} {\bibfnamefont {K.}~\bibnamefont {Gregory}}, \bibinfo {author} {\bibfnamefont {C.~A.}\ \bibnamefont {Puliafito}},  \emph {et~al.},\ }\href@noop {} {\bibfield  {journal} {\bibinfo  {journal} {science}\ }\textbf {\bibinfo {volume} {254}},\ \bibinfo {pages} {1178} (\bibinfo {year} {1991})}\BibitemShut {NoStop}%
\bibitem [{\citenamefont {Stelzer}\ \emph {et~al.}(2021)\citenamefont {Stelzer}, \citenamefont {Strobl}, \citenamefont {Chang}, \citenamefont {Preusser}, \citenamefont {Preibisch}, \citenamefont {McDole},\ and\ \citenamefont {Fiolka}}]{stelzer2021light}%
  \BibitemOpen
  \bibfield  {author} {\bibinfo {author} {\bibfnamefont {E.~H.}\ \bibnamefont {Stelzer}}, \bibinfo {author} {\bibfnamefont {F.}~\bibnamefont {Strobl}}, \bibinfo {author} {\bibfnamefont {B.-J.}\ \bibnamefont {Chang}}, \bibinfo {author} {\bibfnamefont {F.}~\bibnamefont {Preusser}}, \bibinfo {author} {\bibfnamefont {S.}~\bibnamefont {Preibisch}}, \bibinfo {author} {\bibfnamefont {K.}~\bibnamefont {McDole}}, \ and\ \bibinfo {author} {\bibfnamefont {R.}~\bibnamefont {Fiolka}},\ }\href@noop {} {\bibfield  {journal} {\bibinfo  {journal} {Nature Reviews Methods Primers}\ }\textbf {\bibinfo {volume} {1}},\ \bibinfo {pages} {73} (\bibinfo {year} {2021})}\BibitemShut {NoStop}%
\bibitem [{\citenamefont {Wilson}\ \emph {et~al.}(1990)\citenamefont {Wilson} \emph {et~al.}}]{wilson1990confocal}%
  \BibitemOpen
  \bibfield  {author} {\bibinfo {author} {\bibfnamefont {T.}~\bibnamefont {Wilson}} \emph {et~al.},\ }\href@noop {} {\emph {\bibinfo {title} {Confocal microscopy}}},\ Vol.\ \bibinfo {volume} {426}\ (\bibinfo  {publisher} {Academic press London},\ \bibinfo {year} {1990})\BibitemShut {NoStop}%
\bibitem [{\citenamefont {Levoy}\ \emph {et~al.}(2006)\citenamefont {Levoy}, \citenamefont {Ng}, \citenamefont {Adams}, \citenamefont {Footer},\ and\ \citenamefont {Horowitz}}]{levoy2006light}%
  \BibitemOpen
  \bibfield  {author} {\bibinfo {author} {\bibfnamefont {M.}~\bibnamefont {Levoy}}, \bibinfo {author} {\bibfnamefont {R.}~\bibnamefont {Ng}}, \bibinfo {author} {\bibfnamefont {A.}~\bibnamefont {Adams}}, \bibinfo {author} {\bibfnamefont {M.}~\bibnamefont {Footer}}, \ and\ \bibinfo {author} {\bibfnamefont {M.}~\bibnamefont {Horowitz}},\ }in\ \href@noop {} {\emph {\bibinfo {booktitle} {Acm siggraph 2006 papers}}}\ (\bibinfo {year} {2006})\ pp.\ \bibinfo {pages} {924--934}\BibitemShut {NoStop}%
\bibitem [{\citenamefont {Drexler}\ and\ \citenamefont {Fujimoto}(2008)}]{drexler2008optical}%
  \BibitemOpen
  \bibfield  {author} {\bibinfo {author} {\bibfnamefont {W.}~\bibnamefont {Drexler}}\ and\ \bibinfo {author} {\bibfnamefont {J.~G.}\ \bibnamefont {Fujimoto}},\ }\href@noop {} {\emph {\bibinfo {title} {Optical coherence tomography: technology and applications}}}\ (\bibinfo  {publisher} {Springer Science \& Business Media},\ \bibinfo {year} {2008})\BibitemShut {NoStop}%
\bibitem [{\citenamefont {Beaurepaire}\ \emph {et~al.}(1998)\citenamefont {Beaurepaire}, \citenamefont {Boccara}, \citenamefont {Lebec}, \citenamefont {Blanchot},\ and\ \citenamefont {Saint-Jalmes}}]{beaurepaire1998full}%
  \BibitemOpen
  \bibfield  {author} {\bibinfo {author} {\bibfnamefont {E.}~\bibnamefont {Beaurepaire}}, \bibinfo {author} {\bibfnamefont {A.~C.}\ \bibnamefont {Boccara}}, \bibinfo {author} {\bibfnamefont {M.}~\bibnamefont {Lebec}}, \bibinfo {author} {\bibfnamefont {L.}~\bibnamefont {Blanchot}}, \ and\ \bibinfo {author} {\bibfnamefont {H.}~\bibnamefont {Saint-Jalmes}},\ }\href@noop {} {\bibfield  {journal} {\bibinfo  {journal} {Optics letters}\ }\textbf {\bibinfo {volume} {23}},\ \bibinfo {pages} {244} (\bibinfo {year} {1998})}\BibitemShut {NoStop}%
\bibitem [{\citenamefont {Dubois}\ \emph {et~al.}(2002)\citenamefont {Dubois}, \citenamefont {Vabre}, \citenamefont {Boccara},\ and\ \citenamefont {Beaurepaire}}]{dubois2002high}%
  \BibitemOpen
  \bibfield  {author} {\bibinfo {author} {\bibfnamefont {A.}~\bibnamefont {Dubois}}, \bibinfo {author} {\bibfnamefont {L.}~\bibnamefont {Vabre}}, \bibinfo {author} {\bibfnamefont {A.-C.}\ \bibnamefont {Boccara}}, \ and\ \bibinfo {author} {\bibfnamefont {E.}~\bibnamefont {Beaurepaire}},\ }\href@noop {} {\bibfield  {journal} {\bibinfo  {journal} {Applied optics}\ }\textbf {\bibinfo {volume} {41}},\ \bibinfo {pages} {805} (\bibinfo {year} {2002})}\BibitemShut {NoStop}%
\bibitem [{\citenamefont {Apelian}\ \emph {et~al.}(2016)\citenamefont {Apelian}, \citenamefont {Harms}, \citenamefont {Thouvenin},\ and\ \citenamefont {Boccara}}]{apelian_dynamic_2016}%
  \BibitemOpen
  \bibfield  {author} {\bibinfo {author} {\bibfnamefont {C.}~\bibnamefont {Apelian}}, \bibinfo {author} {\bibfnamefont {F.}~\bibnamefont {Harms}}, \bibinfo {author} {\bibfnamefont {O.}~\bibnamefont {Thouvenin}}, \ and\ \bibinfo {author} {\bibfnamefont {A.~C.}\ \bibnamefont {Boccara}},\ }\href {\doibase 10.1364/boe.7.001511} {\bibfield  {journal} {\bibinfo  {journal} {Biomedical Optics Express}\ }\textbf {\bibinfo {volume} {7}},\ \bibinfo {pages} {1511} (\bibinfo {year} {2016})},\ \bibinfo {note} {publisher: Optica Publishing Group}\BibitemShut {NoStop}%
\bibitem [{\citenamefont {Latrive}\ and\ \citenamefont {Boccara}(2011)}]{latrive2011vivo}%
  \BibitemOpen
  \bibfield  {author} {\bibinfo {author} {\bibfnamefont {A.}~\bibnamefont {Latrive}}\ and\ \bibinfo {author} {\bibfnamefont {A.~C.}\ \bibnamefont {Boccara}},\ }\href@noop {} {\bibfield  {journal} {\bibinfo  {journal} {Biomedical optics express}\ }\textbf {\bibinfo {volume} {2}},\ \bibinfo {pages} {2897} (\bibinfo {year} {2011})}\BibitemShut {NoStop}%
\bibitem [{\citenamefont {Scholler}\ \emph {et~al.}(2019)\citenamefont {Scholler}, \citenamefont {Mazlin}, \citenamefont {Thouvenin}, \citenamefont {Groux}, \citenamefont {Xiao}, \citenamefont {Sahel}, \citenamefont {Fink}, \citenamefont {Boccara},\ and\ \citenamefont {Grieve}}]{scholler_probing_2019}%
  \BibitemOpen
  \bibfield  {author} {\bibinfo {author} {\bibfnamefont {J.}~\bibnamefont {Scholler}}, \bibinfo {author} {\bibfnamefont {V.}~\bibnamefont {Mazlin}}, \bibinfo {author} {\bibfnamefont {O.}~\bibnamefont {Thouvenin}}, \bibinfo {author} {\bibfnamefont {K.}~\bibnamefont {Groux}}, \bibinfo {author} {\bibfnamefont {P.}~\bibnamefont {Xiao}}, \bibinfo {author} {\bibfnamefont {J.-A.}\ \bibnamefont {Sahel}}, \bibinfo {author} {\bibfnamefont {M.}~\bibnamefont {Fink}}, \bibinfo {author} {\bibfnamefont {C.}~\bibnamefont {Boccara}}, \ and\ \bibinfo {author} {\bibfnamefont {K.}~\bibnamefont {Grieve}},\ }\href {\doibase 10.1364/BOE.10.000731} {\bibfield  {journal} {\bibinfo  {journal} {Biomedical Optics Express}\ }\textbf {\bibinfo {volume} {10}},\ \bibinfo {pages} {731} (\bibinfo {year} {2019})}\BibitemShut {NoStop}%
\bibitem [{\citenamefont {Morawiec}\ \emph {et~al.}(2024{\natexlab{a}})\citenamefont {Morawiec}, \citenamefont {Ajduk}, \citenamefont {Stremplewski}, \citenamefont {Kennedy},\ and\ \citenamefont {Szkulmowski}}]{morawiec_full-field_2024}%
  \BibitemOpen
  \bibfield  {author} {\bibinfo {author} {\bibfnamefont {S.}~\bibnamefont {Morawiec}}, \bibinfo {author} {\bibfnamefont {A.}~\bibnamefont {Ajduk}}, \bibinfo {author} {\bibfnamefont {P.}~\bibnamefont {Stremplewski}}, \bibinfo {author} {\bibfnamefont {B.~F.}\ \bibnamefont {Kennedy}}, \ and\ \bibinfo {author} {\bibfnamefont {M.}~\bibnamefont {Szkulmowski}},\ }\href {\doibase 10.1038/s42003-024-06745-x} {\bibfield  {journal} {\bibinfo  {journal} {Communications Biology}\ }\textbf {\bibinfo {volume} {7}},\ \bibinfo {pages} {1057} (\bibinfo {year} {2024}{\natexlab{a}})}\BibitemShut {NoStop}%
\bibitem [{\citenamefont {Scholler}\ \emph {et~al.}(2020)\citenamefont {Scholler}, \citenamefont {Groux}, \citenamefont {Goureau}, \citenamefont {Sahel}, \citenamefont {Fink}, \citenamefont {Reichman}, \citenamefont {Boccara},\ and\ \citenamefont {Grieve}}]{scholler_dynamic_2020}%
  \BibitemOpen
  \bibfield  {author} {\bibinfo {author} {\bibfnamefont {J.}~\bibnamefont {Scholler}}, \bibinfo {author} {\bibfnamefont {K.}~\bibnamefont {Groux}}, \bibinfo {author} {\bibfnamefont {O.}~\bibnamefont {Goureau}}, \bibinfo {author} {\bibfnamefont {J.-A.}\ \bibnamefont {Sahel}}, \bibinfo {author} {\bibfnamefont {M.}~\bibnamefont {Fink}}, \bibinfo {author} {\bibfnamefont {S.}~\bibnamefont {Reichman}}, \bibinfo {author} {\bibfnamefont {C.}~\bibnamefont {Boccara}}, \ and\ \bibinfo {author} {\bibfnamefont {K.}~\bibnamefont {Grieve}},\ }\href {\doibase 10.1038/s41377-020-00375-8} {\bibfield  {journal} {\bibinfo  {journal} {Light: Science \& Applications}\ }\textbf {\bibinfo {volume} {9}},\ \bibinfo {pages} {140} (\bibinfo {year} {2020})}\BibitemShut {NoStop}%
\bibitem [{\citenamefont {Monfort}\ \emph {et~al.}(2025)\citenamefont {Monfort}, \citenamefont {Grieve},\ and\ \citenamefont {Thouvenin}}]{monfort2025rolling}%
  \BibitemOpen
  \bibfield  {author} {\bibinfo {author} {\bibfnamefont {T.}~\bibnamefont {Monfort}}, \bibinfo {author} {\bibfnamefont {K.}~\bibnamefont {Grieve}}, \ and\ \bibinfo {author} {\bibfnamefont {O.}~\bibnamefont {Thouvenin}},\ }\href@noop {} {\bibfield  {journal} {\bibinfo  {journal} {Optics Letters}\ }\textbf {\bibinfo {volume} {50}},\ \bibinfo {pages} {2239} (\bibinfo {year} {2025})}\BibitemShut {NoStop}%
\bibitem [{\citenamefont {Recher}\ \emph {et~al.}(2020)\citenamefont {Recher}, \citenamefont {Nassoy},\ and\ \citenamefont {Badon}}]{recher2020remote}%
  \BibitemOpen
  \bibfield  {author} {\bibinfo {author} {\bibfnamefont {G.}~\bibnamefont {Recher}}, \bibinfo {author} {\bibfnamefont {P.}~\bibnamefont {Nassoy}}, \ and\ \bibinfo {author} {\bibfnamefont {A.}~\bibnamefont {Badon}},\ }\href@noop {} {\bibfield  {journal} {\bibinfo  {journal} {Biomedical optics express}\ }\textbf {\bibinfo {volume} {11}},\ \bibinfo {pages} {2578} (\bibinfo {year} {2020})}\BibitemShut {NoStop}%
\bibitem [{\citenamefont {Auksorius}\ \emph {et~al.}(2022)\citenamefont {Auksorius}, \citenamefont {Borycki}, \citenamefont {Wegrzyn}, \citenamefont {Sikorski}, \citenamefont {Lizewski}, \citenamefont {Zickiene}, \citenamefont {Rapolu}, \citenamefont {Adomavicius}, \citenamefont {Tomczewski},\ and\ \citenamefont {Wojtkowski}}]{auksorius2022spatio}%
  \BibitemOpen
  \bibfield  {author} {\bibinfo {author} {\bibfnamefont {E.}~\bibnamefont {Auksorius}}, \bibinfo {author} {\bibfnamefont {D.}~\bibnamefont {Borycki}}, \bibinfo {author} {\bibfnamefont {P.}~\bibnamefont {Wegrzyn}}, \bibinfo {author} {\bibfnamefont {B.~L.}\ \bibnamefont {Sikorski}}, \bibinfo {author} {\bibfnamefont {K.}~\bibnamefont {Lizewski}}, \bibinfo {author} {\bibfnamefont {I.}~\bibnamefont {Zickiene}}, \bibinfo {author} {\bibfnamefont {M.}~\bibnamefont {Rapolu}}, \bibinfo {author} {\bibfnamefont {K.}~\bibnamefont {Adomavicius}}, \bibinfo {author} {\bibfnamefont {S.}~\bibnamefont {Tomczewski}}, \ and\ \bibinfo {author} {\bibfnamefont {M.}~\bibnamefont {Wojtkowski}},\ }\href@noop {} {\bibfield  {journal} {\bibinfo  {journal} {IScience}\ }\textbf {\bibinfo {volume} {25}} (\bibinfo {year} {2022})}\BibitemShut {NoStop}%
\bibitem [{\citenamefont {Alessandri}\ \emph {et~al.}(2013)\citenamefont {Alessandri}, \citenamefont {Sarangi}, \citenamefont {Gurchenkov}, \citenamefont {Sinha}, \citenamefont {Kießling}, \citenamefont {Fetler}, \citenamefont {Rico}, \citenamefont {Scheuring}, \citenamefont {Lamaze}, \citenamefont {Simon}, \citenamefont {Geraldo}, \citenamefont {Vignjević}, \citenamefont {Doméjean}, \citenamefont {Rolland}, \citenamefont {Funfak}, \citenamefont {Bibette}, \citenamefont {Bremond},\ and\ \citenamefont {Nassoy}}]{alessandri_cellular_2013}%
  \BibitemOpen
  \bibfield  {author} {\bibinfo {author} {\bibfnamefont {K.}~\bibnamefont {Alessandri}}, \bibinfo {author} {\bibfnamefont {B.~R.}\ \bibnamefont {Sarangi}}, \bibinfo {author} {\bibfnamefont {V.~V.}\ \bibnamefont {Gurchenkov}}, \bibinfo {author} {\bibfnamefont {B.}~\bibnamefont {Sinha}}, \bibinfo {author} {\bibfnamefont {T.~R.}\ \bibnamefont {Kießling}}, \bibinfo {author} {\bibfnamefont {L.}~\bibnamefont {Fetler}}, \bibinfo {author} {\bibfnamefont {F.}~\bibnamefont {Rico}}, \bibinfo {author} {\bibfnamefont {S.}~\bibnamefont {Scheuring}}, \bibinfo {author} {\bibfnamefont {C.}~\bibnamefont {Lamaze}}, \bibinfo {author} {\bibfnamefont {A.}~\bibnamefont {Simon}}, \bibinfo {author} {\bibfnamefont {S.}~\bibnamefont {Geraldo}}, \bibinfo {author} {\bibfnamefont {D.}~\bibnamefont {Vignjević}}, \bibinfo {author} {\bibfnamefont {H.}~\bibnamefont {Doméjean}}, \bibinfo {author} {\bibfnamefont {L.}~\bibnamefont {Rolland}}, \bibinfo {author} {\bibfnamefont {A.}~\bibnamefont {Funfak}}, \bibinfo {author} {\bibfnamefont
  {J.}~\bibnamefont {Bibette}}, \bibinfo {author} {\bibfnamefont {N.}~\bibnamefont {Bremond}}, \ and\ \bibinfo {author} {\bibfnamefont {P.}~\bibnamefont {Nassoy}},\ }\href {\doibase 10.1073/pnas.1309482110} {\bibfield  {journal} {\bibinfo  {journal} {Proceedings of the National Academy of Sciences}\ }\textbf {\bibinfo {volume} {110}},\ \bibinfo {pages} {14843} (\bibinfo {year} {2013})},\ \bibinfo {note} {publisher: Proceedings of the National Academy of Sciences}\BibitemShut {NoStop}%
\bibitem [{\citenamefont {Cohen}\ \emph {et~al.}(2023)\citenamefont {Cohen}, \citenamefont {Luquet}, \citenamefont {Pletenka}, \citenamefont {Leonard}, \citenamefont {Warter}, \citenamefont {Gurchenkov}, \citenamefont {Carrere}, \citenamefont {Rieu}, \citenamefont {Hardouin}, \citenamefont {Moncaubeig} \emph {et~al.}}]{cohen2023engineering}%
  \BibitemOpen
  \bibfield  {author} {\bibinfo {author} {\bibfnamefont {P.~J.}\ \bibnamefont {Cohen}}, \bibinfo {author} {\bibfnamefont {E.}~\bibnamefont {Luquet}}, \bibinfo {author} {\bibfnamefont {J.}~\bibnamefont {Pletenka}}, \bibinfo {author} {\bibfnamefont {A.}~\bibnamefont {Leonard}}, \bibinfo {author} {\bibfnamefont {E.}~\bibnamefont {Warter}}, \bibinfo {author} {\bibfnamefont {B.}~\bibnamefont {Gurchenkov}}, \bibinfo {author} {\bibfnamefont {J.}~\bibnamefont {Carrere}}, \bibinfo {author} {\bibfnamefont {C.}~\bibnamefont {Rieu}}, \bibinfo {author} {\bibfnamefont {J.}~\bibnamefont {Hardouin}}, \bibinfo {author} {\bibfnamefont {F.}~\bibnamefont {Moncaubeig}},  \emph {et~al.},\ }\href@noop {} {\bibfield  {journal} {\bibinfo  {journal} {Biomaterials}\ }\textbf {\bibinfo {volume} {295}},\ \bibinfo {pages} {122033} (\bibinfo {year} {2023})}\BibitemShut {NoStop}%
\bibitem [{\citenamefont {Suire}\ \emph {et~al.}(2025)\citenamefont {Suire}, \citenamefont {Jana}, \citenamefont {Nassoy},\ and\ \citenamefont {Badon}}]{suire2025extending}%
  \BibitemOpen
  \bibfield  {author} {\bibinfo {author} {\bibfnamefont {L.}~\bibnamefont {Suire}}, \bibinfo {author} {\bibfnamefont {A.}~\bibnamefont {Jana}}, \bibinfo {author} {\bibfnamefont {P.}~\bibnamefont {Nassoy}}, \ and\ \bibinfo {author} {\bibfnamefont {A.}~\bibnamefont {Badon}},\ }\href@noop {} {\bibfield  {journal} {\bibinfo  {journal} {Advanced Science}\ ,\ \bibinfo {pages} {e05501}} (\bibinfo {year} {2025})}\BibitemShut {NoStop}%
\bibitem [{\citenamefont {Baczewska}\ \emph {et~al.}(2021)\citenamefont {Baczewska}, \citenamefont {Eder}, \citenamefont {Ketelhut}, \citenamefont {Kemper},\ and\ \citenamefont {Kujawińska}}]{baczewska_refractive_2021}%
  \BibitemOpen
  \bibfield  {author} {\bibinfo {author} {\bibfnamefont {M.}~\bibnamefont {Baczewska}}, \bibinfo {author} {\bibfnamefont {K.}~\bibnamefont {Eder}}, \bibinfo {author} {\bibfnamefont {S.}~\bibnamefont {Ketelhut}}, \bibinfo {author} {\bibfnamefont {B.}~\bibnamefont {Kemper}}, \ and\ \bibinfo {author} {\bibfnamefont {M.}~\bibnamefont {Kujawińska}},\ }\href {\doibase 10.1002/cyto.a.24229} {\bibfield  {journal} {\bibinfo  {journal} {Cytometry Part A}\ }\textbf {\bibinfo {volume} {99}},\ \bibinfo {pages} {388} (\bibinfo {year} {2021})}\BibitemShut {NoStop}%
\bibitem [{\citenamefont {Morawiec}\ \emph {et~al.}(2024{\natexlab{b}})\citenamefont {Morawiec}, \citenamefont {Ajduk}, \citenamefont {Stremplewski}, \citenamefont {Kennedy},\ and\ \citenamefont {Szkulmowski}}]{morawiec2024full}%
  \BibitemOpen
  \bibfield  {author} {\bibinfo {author} {\bibfnamefont {S.}~\bibnamefont {Morawiec}}, \bibinfo {author} {\bibfnamefont {A.}~\bibnamefont {Ajduk}}, \bibinfo {author} {\bibfnamefont {P.}~\bibnamefont {Stremplewski}}, \bibinfo {author} {\bibfnamefont {B.~F.}\ \bibnamefont {Kennedy}}, \ and\ \bibinfo {author} {\bibfnamefont {M.}~\bibnamefont {Szkulmowski}},\ }\href@noop {} {\bibfield  {journal} {\bibinfo  {journal} {Communications Biology}\ }\textbf {\bibinfo {volume} {7}},\ \bibinfo {pages} {1057} (\bibinfo {year} {2024}{\natexlab{b}})}\BibitemShut {NoStop}%
\bibitem [{\citenamefont {Yin}\ \emph {et~al.}(2025)\citenamefont {Yin}, \citenamefont {He}, \citenamefont {Ying}, \citenamefont {Zhang}, \citenamefont {Yang}, \citenamefont {Chen}, \citenamefont {Hu}, \citenamefont {Shi}, \citenamefont {Xue}, \citenamefont {Wang}, \citenamefont {Wang}, \citenamefont {Wang},\ and\ \citenamefont {Xue}}]{yin_fast_2025}%
  \BibitemOpen
  \bibfield  {author} {\bibinfo {author} {\bibfnamefont {Z.}~\bibnamefont {Yin}}, \bibinfo {author} {\bibfnamefont {B.}~\bibnamefont {He}}, \bibinfo {author} {\bibfnamefont {Y.}~\bibnamefont {Ying}}, \bibinfo {author} {\bibfnamefont {S.}~\bibnamefont {Zhang}}, \bibinfo {author} {\bibfnamefont {P.}~\bibnamefont {Yang}}, \bibinfo {author} {\bibfnamefont {Z.}~\bibnamefont {Chen}}, \bibinfo {author} {\bibfnamefont {Z.}~\bibnamefont {Hu}}, \bibinfo {author} {\bibfnamefont {Y.}~\bibnamefont {Shi}}, \bibinfo {author} {\bibfnamefont {R.}~\bibnamefont {Xue}}, \bibinfo {author} {\bibfnamefont {C.}~\bibnamefont {Wang}}, \bibinfo {author} {\bibfnamefont {S.}~\bibnamefont {Wang}}, \bibinfo {author} {\bibfnamefont {G.}~\bibnamefont {Wang}}, \ and\ \bibinfo {author} {\bibfnamefont {P.}~\bibnamefont {Xue}},\ }\href {\doibase 10.1038/s44303-025-00068-0} {\bibfield  {journal} {\bibinfo  {journal} {npj Imaging}\ }\textbf {\bibinfo {volume} {3}},\ \bibinfo {pages} {12} (\bibinfo {year} {2025})}\BibitemShut {NoStop}%
\bibitem [{\citenamefont {Badon}\ \emph {et~al.}(2022)\citenamefont {Badon}, \citenamefont {Andrique}, \citenamefont {Mombereau}, \citenamefont {Rivet}, \citenamefont {Boyreau}, \citenamefont {Nassoy},\ and\ \citenamefont {Recher}}]{badon2022incubascope}%
  \BibitemOpen
  \bibfield  {author} {\bibinfo {author} {\bibfnamefont {A.}~\bibnamefont {Badon}}, \bibinfo {author} {\bibfnamefont {L.}~\bibnamefont {Andrique}}, \bibinfo {author} {\bibfnamefont {A.}~\bibnamefont {Mombereau}}, \bibinfo {author} {\bibfnamefont {L.}~\bibnamefont {Rivet}}, \bibinfo {author} {\bibfnamefont {A.}~\bibnamefont {Boyreau}}, \bibinfo {author} {\bibfnamefont {P.}~\bibnamefont {Nassoy}}, \ and\ \bibinfo {author} {\bibfnamefont {G.}~\bibnamefont {Recher}},\ }\href@noop {} {\bibfield  {journal} {\bibinfo  {journal} {Royal Society Open Science}\ }\textbf {\bibinfo {volume} {9}},\ \bibinfo {pages} {211444} (\bibinfo {year} {2022})}\BibitemShut {NoStop}%
\bibitem [{\citenamefont {Jana}\ \emph {et~al.}(2025)\citenamefont {Jana}, \citenamefont {Mekhileri}, \citenamefont {Boyreau}, \citenamefont {Bazin}, \citenamefont {Pujol}, \citenamefont {Alessandri}, \citenamefont {Recher}, \citenamefont {Nassoy},\ and\ \citenamefont {Badon}}]{jana2025zincubascope}%
  \BibitemOpen
  \bibfield  {author} {\bibinfo {author} {\bibfnamefont {A.}~\bibnamefont {Jana}}, \bibinfo {author} {\bibfnamefont {N.}~\bibnamefont {Mekhileri}}, \bibinfo {author} {\bibfnamefont {A.}~\bibnamefont {Boyreau}}, \bibinfo {author} {\bibfnamefont {A.}~\bibnamefont {Bazin}}, \bibinfo {author} {\bibfnamefont {N.}~\bibnamefont {Pujol}}, \bibinfo {author} {\bibfnamefont {K.}~\bibnamefont {Alessandri}}, \bibinfo {author} {\bibfnamefont {G.}~\bibnamefont {Recher}}, \bibinfo {author} {\bibfnamefont {P.}~\bibnamefont {Nassoy}}, \ and\ \bibinfo {author} {\bibfnamefont {A.}~\bibnamefont {Badon}},\ }\href@noop {} {\bibfield  {journal} {\bibinfo  {journal} {PloS one}\ }\textbf {\bibinfo {volume} {20}},\ \bibinfo {pages} {e0309035} (\bibinfo {year} {2025})}\BibitemShut {NoStop}%
\end{thebibliography}%
